\documentclass[aps,prl,twocolumn,superscriptaddress]{revtex4}
\usepackage{graphicx,epsf}
\usepackage{amsmath,amssymb}

\usepackage{color}

\newcommand{\wvq}{\omega_{\vec q}}
\newcommand{\wk}{\omega_k}

\newcommand{\gvq}{\gamma_{\vec q}}

\newcommand{\HSW}{\mathcal{H}_{\rm SW}}

\newcommand{\dis}{U} 
\newcommand{\uu}{u}  

\newcommand{\Cmax}{C_{\rm max}}

\begin{document}

\title{
Magnon Landau levels and emergent supersymmetry in strained antiferromagnets
}

\author{Mary Madelynn Nayga}
\affiliation{Institut f\"ur Theoretische Physik and W\"urzburg-Dresden Cluster of Excellence ct.qmat, Technische Universit\"at Dresden,
01062 Dresden, Germany}
\affiliation{Max-Planck-Institut f\"ur Chemische Physik fester Stoffe, N\"othnitzer Str. 40,
01187 Dresden, Germany}
\author{Stephan Rachel}
\affiliation{School of Physics, University of Melbourne, Parkville, VIC 3010, Australia}
\author{Matthias Vojta}
\affiliation{Institut f\"ur Theoretische Physik and W\"urzburg-Dresden Cluster of Excellence ct.qmat, Technische Universit\"at Dresden,
01062 Dresden, Germany}

\date{\today}


\begin{abstract}
Inhomogeneous strain applied to lattice systems can induce artificial gauge fields for particles moving on this lattice. Here we demonstrate how to engineer a novel state of matter, namely an antiferromagnet with a Landau-level excitation spectrum of magnons. We consider a honeycomb-lattice Heisenberg model and show that triaxial strain leads to equally spaced pseudo-Landau levels at the upper end of the magnon spectrum, with degeneracies characteristic of emergent supersymmetry. We also present a particular strain protocol which induces perfectly quantized magnon Landau levels over the whole bandwidth. We discuss experimental realizations and generalizations.
\end{abstract}
\maketitle


Artificial gauge fields have become a powerful tool to engineer states of matter. Originally discussed in the context of strain applied to carbon nanotubes \cite{suzuura02}, they have later been utilized to induce Landau levels in strained graphene \cite{guinea09,levy10,strain-arc,voz_rev10}, with a spacing corresponding to a magnetic field exceeding 300\,T, far beyond what is reachable with  laboratory magnetic fields.
More fundamentally, artificial gauge fields can be used to induce effects akin to that of orbital magnetic fields for \textit{charge-neutral} particles, resulting in novel states which cannot be generated otherwise. This has been proposed for ultracold gases as well as solids \cite{cold_gauge,aidelsburger18}, with strain-induced Landau levels for Bogoliubov quasiparticles of nodal superconductors \cite{wachtel17,franz18} and for Majorana excitations of a Kitaev spin liquid \cite{rachel16a} being two prominent examples.

In this Letter, we propose a novel setting for strain engineering, namely ordered quantum antiferromagnets. We show that inhomogeneous exchange couplings in simple N\'eel antiferromagnets, generated by applying a suitable strain pattern, transform the conventional magnon spectrum into a sequence of magnonic pseudo-Landau levels. There are three key differences to previous condensed-matter realizations of pseudo-Landau levels:
(i) The magnon Landau levels emerge in the high-energy part of the spectrum, i.e., starting from the upper band edge, not at low energies as in the graphene case.
(ii) The magnon Landau levels derive from a mode which disperses quadratically in the absence of strain, not from a linear Dirac-like dispersion. As a result, the magnon Landau levels are equally spaced, with a spacing scaling linearly with the pseudo-magnetic field, as opposed to the square-root dependence of Dirac Landau levels.
(iii) Perhaps most remarkably, the Landau-level spectrum displays emergent supersymmetry: Its degeneracies are equivalent to that of a system with a boson and a fermion of equal energy, thus forming one of the rare condensed-matter realizations of supersymmetry \cite{kulish80,grover14,jian17}, which in general refers to systems where each boson has a fermionic superpartner and vice versa.

We present numerical spin-wave results for the Heisenberg antiferromagnet on finite honeycomb lattices subject to triaxial strain, and we derive the corresponding continuum field theory which is based on an expansion about the \emph{maximum} of the magnon dispersion. Following earlier work \cite{rachel16b}, we also show that the combined limit of strong magneto-elastic coupling and weak lattice deformations leads to perfectly quantized magnon Landau levels over the entire range of magnon energies.
We argue that magnon Landau levels can be detected in high-resolution spectroscopic experiments on strained honeycomb antiferromagnets, and we discuss generalizations to other lattices.

We note that very different types of magnon Landau levels have appeared in earlier work: Ref.~\onlinecite{nakata17} considered gapped antiferromagnets subject to electric field gradients, and Ref.~\onlinecite{ferreiros18} studied Dirac magnons in ferromagnets under strain.


\textit{Model and spin-wave theory.}
We consider a standard nearest-neighbor antiferromagnetic Heisenberg model of spins $S$ placed on the sites of a honeycomb lattice. The Hamiltonian with spatially varying couplings reads
\begin{equation}
\label{hk}
\mathcal{H} = \sum_{\langle ij\rangle} J_{ij} \vec{S}_i \cdot \vec{S}_j\,.
\end{equation}
On this bipartite lattice, the antiferromagnetic couplings are unfrustrated, such that the ground state of the homogeneous system, $J_{ij}\equiv J$, is a simple N\'eel antiferromagnet for any $S$. %
In the following we will assume that a N\'eel state is also realized in the strained system; this applies in the semiclassical ($S\to\infty$) limit \cite{classnote} as well as, by continuity, for weak strain and any $S$.

The excitation spectrum can be obtained using spin-wave theory. Using Holstein-Primakoff bosons $a$ and $b$ on sublattices $A$ and $B$, respectively, which describe fluctuations above the collinear N\'eel state, the bilinear piece of the Hamiltonian reads
\begin{equation}
\label{hsw}
\HSW = S \sum_{\langle ij\rangle} J_{ij} \left(a_i^\dagger a_i + b_j^\dagger b_j + a_i b_j + a_i^\dagger b_j^\dagger \right).
\end{equation}
In the homogeneous (i.e., unstrained) case, the Hamiltonian can be transformed by subsequent Fourier and Bogoliubov transformations into a system of non-interacting bosons with energies
\begin{equation}
\label{specsw}
\wvq = 3JS \sqrt{1 - |\gvq|^2}
\end{equation}
where $\gvq = (1/3) \sum_j e^{i\vec{q}\cdot\vec{\delta_j}}$ with $\vec{\delta}_j$ being the three nearest-neighbor vectors on the honeycomb lattice. This spectrum displays gapless (Goldstone) modes at $\vec{q}=0$ and is maximum at the momenta
$\vec{q} = \vec K= 2\pi/(3a_0)(1/\sqrt{3},1)$ and
$\vec K'= -2\pi/(3a_0)(1/\sqrt{3},1)$,
with $a_0$ the lattice constant.

\begin{figure}[t!]
\includegraphics[width=0.45\columnwidth]{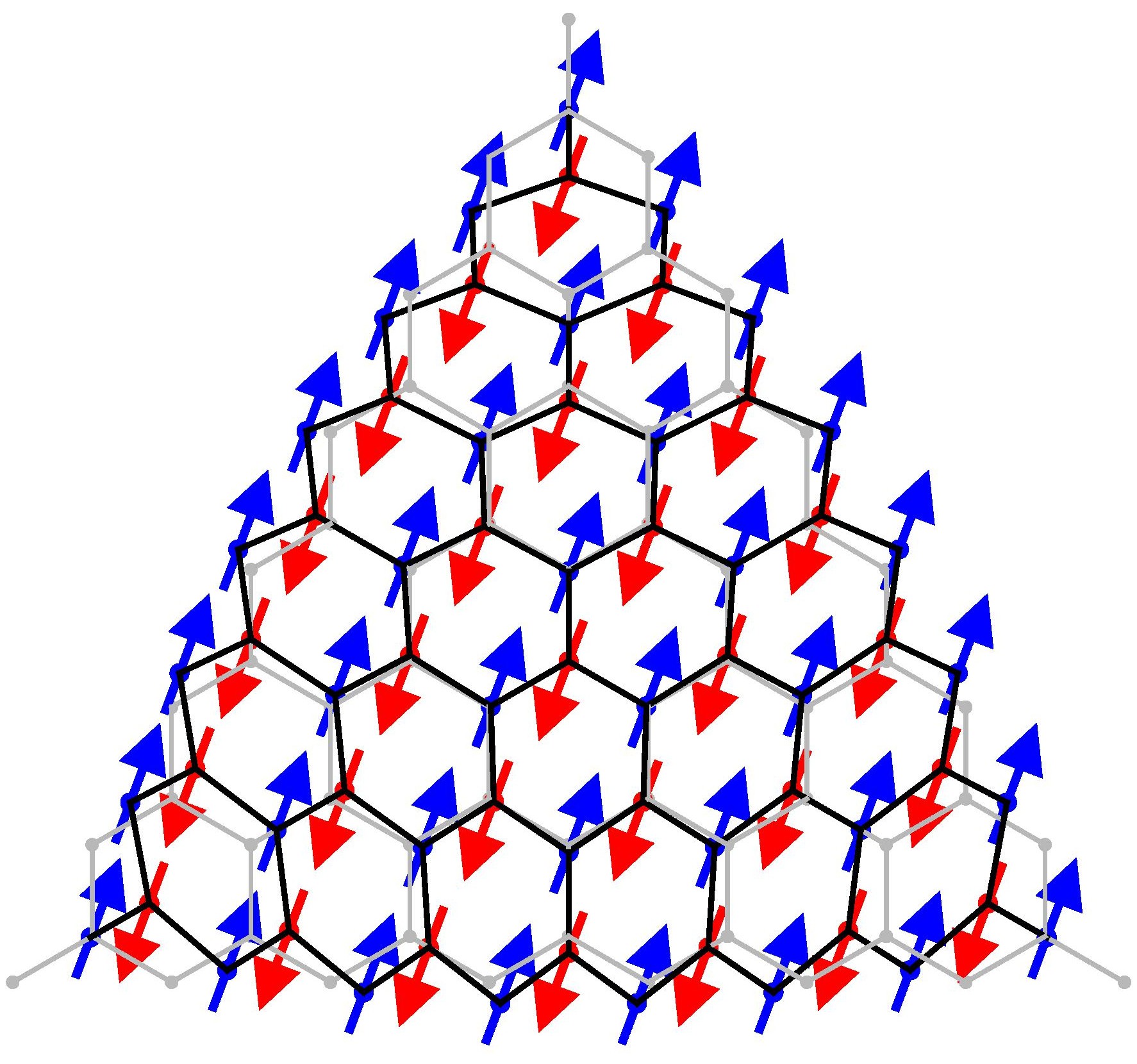}\hfill
\includegraphics[width=0.5\columnwidth]{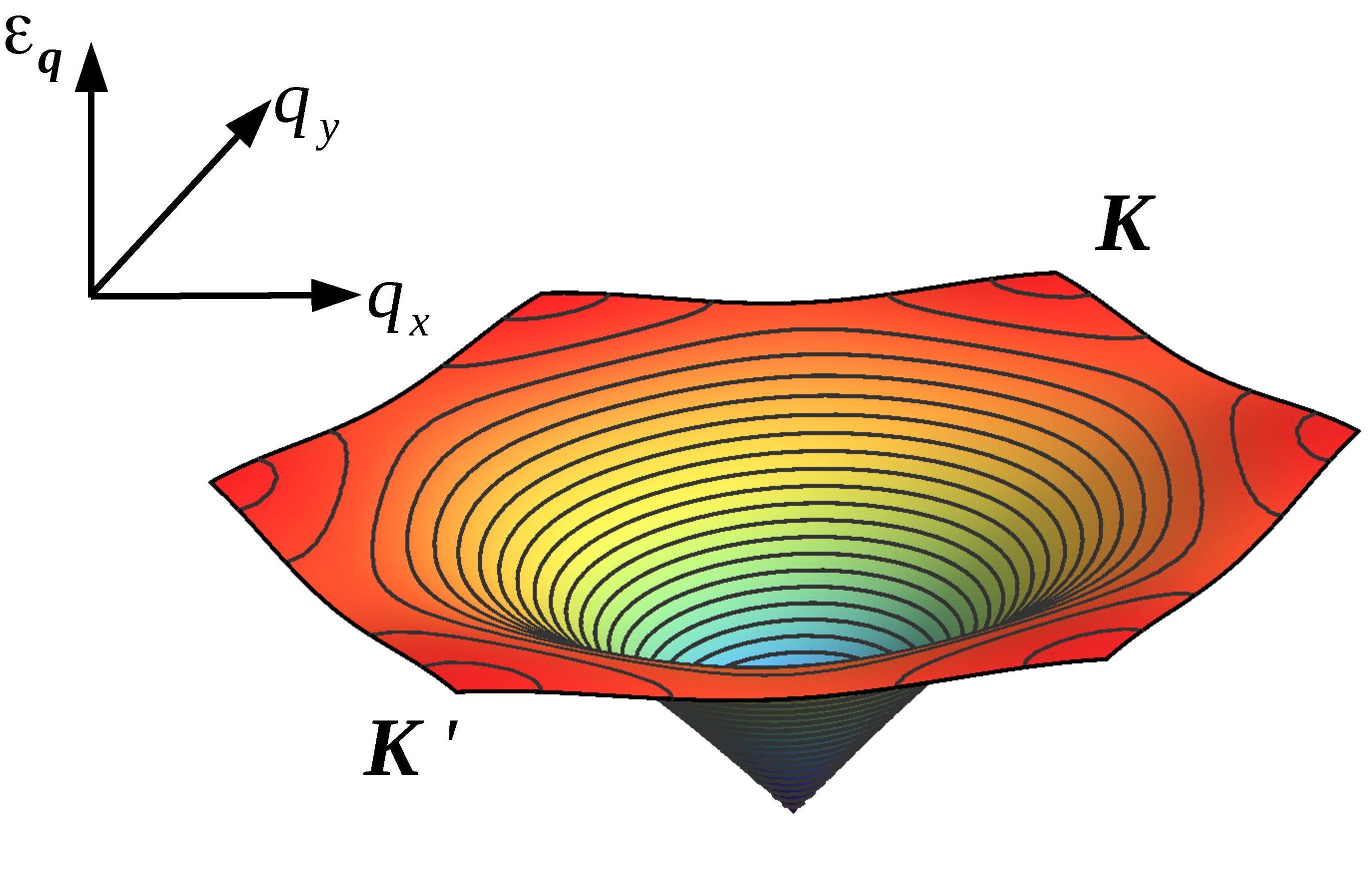}
\caption{
Left:
Distorted honeycomb-lattice antiferromagnet, with displacements from triaxial strain, Eq.~\eqref{displace}; the undistorted lattice is shown in light gray. Longer (shorter) bonds correspond to weaker (stronger) exchange couplings $J_{ij}$. The system size is $N=8$.
Right:
Magnon spectrum of the unstrained honeycomb Heisenberg antiferromagnet, showing a quadratic maximum at momenta $\vec K$ and $\vec K'$.
}
\label{fig:setup}
\end{figure}


\textit{Strain-induced Landau levels.}
In the context of graphene, it has been theoretically shown \cite{guinea09,voz_rev10} that a spatial modulation of hopping energies mimics the effect of a vector potential. Near the Dirac energy, this emergent vector potential can be expressed through the strain tensor $\uu_{ij}$ as
$\vec A \propto \pm ( \uu_{xx} - \uu_{yy} , -2 \uu_{xy} )^T$.
If the resulting pseudo-magnetic field, $\vec B={\rm rot}\vec A$, is sufficiently homogeneous -- this applies, e.g., to triaxial strain with the displacement vector given by \cite{guinea09,peeters13}
\begin{equation}
\label{displace}
\vec{\dis}(x,y) = \bar{C} \big(2xy, x^2-y^2 \big)^T
\end{equation}
-- it can induce single-particle pseudo-Landau levels very similar to Landau levels in a physical magnetic field.
In Eq.~\eqref{displace} $\bar{C}$ (measured in units of $1/a_0$) parameterizes the distortion, and $\uu_{ij} = (\partial_i \dis_j + \partial_j \dis_i)/2$.

Adapting the idea of strain-induced artificial gauge fields to antiferromagnets, we shall study the Heisenberg model \eqref{hk} with spatially modulated exchange couplings.
We choose
\begin{equation}
\label{ourjij}
J_{ij} = J \left[ 1- \beta (|\vec{\delta}_{ij}|/a_0 - 1)\right]
\end{equation}
where we calculate the distance $\vec{\delta}_{ij}=\vec{R}_i+\vec{U}_i-\vec{R}_j-\vec{U}_j$ using the displacement $\vec{U}(x,y)$ evaluated at the lattice positions $\vec{R}_i$ of the undistorted honeycomb lattice. The factor $\beta$ encodes the strength of magneto-elastic coupling, and the dimensionless parameter (dubbed ``strain'' below)
$C = \bar{C}\beta a_0$
will enter our simulations as a measure of the modulations of the $J_{ij}$.
Note that Eq.~\eqref{ourjij} represents a linear approximation to the full dependence of the exchange constant on the bond length \cite{peeters13}.

The inhomogeneous Heisenberg model \eqref{hk} with couplings given by Eq.\,\eqref{ourjij} is expected to display magnon excitations (on top of its collinear ground state) which are influenced by artificial gauge fields. As we demonstrate now, triaxial strain \eqref{displace} will produce a homogeneous pseudo-magnetic field for magnons with momenta near $\vec{K}$ and $\vec{K'}$ leading to equidistant pseudo-Landau levels.

\textit{Field theory for strained magnons.}
We derive a continuum field theory which captures the effect of the strain-induced gauge field. To this end, we expand the magnon Hamiltonian near $\vec{K}$. We introduce the two-vector $\Psi(\vec{r})=(\psi_a(\vec{r}),\psi_b^\dagger(\vec{r}))^T$, representing the real-space Fourier transform of $(a_{\vec{K}+\vec{k}},b_{-\vec{K}+\vec{k}}^\dagger)$ for small $\vec k$.
Working to linear order in the lattice distortion and to linear order in $\vec k$, we find that the magnon Hamiltonian can be written as \cite{suppl}
\begin{equation}
\label{hstrain}
\HSW = 3JS\! \int\! d^2r \Psi^\dagger\!
\begin{pmatrix}
1+a & -\frac{a_0}{2} (\Pi_x \!-\! i \Pi_y) \\
-\frac{a_0}{2} (\Pi_x \!+\! i \Pi_y) & 1+a
\end{pmatrix}
\!\Psi
\end{equation}
where $\vec{\Pi} =  \vec{p} + \vec{A}$ and $p_x=-i\partial_x$, $p_y=-i\partial_y$.
Here, $a = - \beta (\uu_{xx}+\uu_{yy})/2$ encodes the strain-induced change of the magnon bandwidth, and the strain-induced gauge field $\vec{A}$ has the form
\begin{equation}
\label{eqn: pseudo A}
\vec{A} = \dfrac{\beta}{2a_0}
\begin{pmatrix}
\uu_{xx}-\uu_{yy}  \\
-2\uu_{xy}
\end{pmatrix} \,
+ \vec{A}_K \,.
\end{equation}
This consists of a $\beta$-dependent part responsible for Landau-level physics \cite{guinea09,voz_rev10} and a $\beta$-independent lattice correction $\vec{A}_K$ \cite{kitt12,masir13} which will be neglected in the following, for details see Ref.~\onlinecite{suppl}.
In the unstrained case, $\vec A=0$ and $a=0$, the spectrum of the Hamiltonian \eqref{hstrain} can be obtained by Fourier and Bogoliubov transformations as $\wk = 3JS (1- k^2 a_0^2/8)$, consistent with Eq.~\eqref{specsw} for $\vec q=\vec K+\vec k$ and small $\vec k$.

The Hamiltonian \eqref{hstrain} differs in three crucial points from other realizations of strain-induced gauge fields:
(i) It describe bosonic quasiparticles, not fermionic electrons.
(ii) Particle number is not conserved. As a result, a bosonic Bogoliubov transformation is required to diagonalize \eqref{hstrain}, technically different from the diagonalization of a Hermitian matrix.
(iii) The diagonal elements in \eqref{hstrain} are non-zero, as a result of magnon on-site energies.

We now solve the problem explicitly for triaxial strain \eqref{displace} where $a=0$ by symmetry, and the $\beta$-dependent part of the vector potential reads $\vec A = 2 \beta \bar{C} (y/a_0,-x/a_0)^T$ corresponding to symmetric gauge.

We introduce two types of bosonic ladder operators
\begin{align}
\alpha &= \left[p_x + y \frac{2C}{a_0^2} + i (p_y - x \frac{2C}{a_0^2}) \right] \frac{a_0}{\sqrt{8C}} \,,\\
\gamma &= \left[p_x - y \frac{2C}{a_0^2} - i (p_y + x \frac{2C}{a_0^2}) \right] \frac{a_0}{\sqrt{8C}}
\end{align}
which obey $[\alpha,\alpha^\dagger]_-=[\gamma,\gamma^\dagger]_-=1$ and $[\alpha^{(\dagger)},\gamma^{(\dagger)}]_-=0$, and we have assumed $C>0$.
The Hamiltonian now reads
\begin{equation}
\label{hstrain2}
\HSW = 3JS
\int \! d^2r
\Psi^\dagger(\vec r)
\begin{pmatrix}
1 & -{\sqrt{2C}} \alpha^\dagger  \\
-{\sqrt{2C}} \alpha & 1
\end{pmatrix}
\Psi(\vec r) \,.
\end{equation}
As usual for Landau levels in symmetric gauge, the spectrum is determined by the $\alpha$ boson operators, whereas the $\gamma$ boson operators commute with the $\alpha$ and the Hamiltonian, such that all single-particle levels display an extensive degeneracy corresponding to $m = \gamma^\dagger\gamma$.

The single-particle (magnon) eigenstates can be obtained from \eqref{hstrain2} by a bosonic Bogoliubov transformation \cite{wessel05}. The solutions of the corresponding differential equations can be written in real space as \cite{suppl}
\begin{align}
\label{solu}
\Phi_{n,m}^\pm =
\begin{pmatrix}
c^\pm_n \phi_{n,m} \\
\phi_{n-1,m} \\
\end{pmatrix}
,~~
\Phi_{0,m}^+ =
\begin{pmatrix}
\phi_{0,m} \\
0 \\
\end{pmatrix}
,
\end{align}
with integer quantum numbers $n\geq1$ and $m\geq0$. The wavefunctions $\phi_{n,m}$ are harmonic-oscillator eigenstates with $\phi_{n,m} = (n!m!)^{-1/2} (\alpha^\dagger)^n (\gamma^\dagger)^m \phi_0$ with $\alpha \phi_0=\gamma \phi_0=0$, in real space $\phi_0(r) \propto \exp(-Cr^*r/(4a_0^2))$ where $r=x+iy$, and the coefficients are
$c_n^+=\sqrt{2/(Cn)}$ and $c_n^-=\sqrt{Cn/2}$.
Hence, we have a single set of $n=0$ solutions and paired sets of solutions with $n=1, 2, 3, \ldots$
The Hamiltonian \eqref{hstrain2} is diagonalized by magnon operators $A_{n,m}^\pm = \int d^2 r \Psi^T(r) \Phi_{n,m}^\pm(r)$ and takes the Landau-level form
\begin{equation}
\HSW = \sum_m \left( E_0  (A_{0,m}^+)^\dagger A_{0,m}^+ + \sum_{n=1,\pm}^\infty E_n (A_{n,m}^\pm)^\dagger A_{n,m}^\pm \right)
\end{equation}
up to a global constant. The magnon energies are
\begin{equation}
\label{magen}
E_n = 3JS \sqrt{1 - 2 C n} \approx 3JS (1 - C n),~~n=0,1,2,\ldots
\end{equation}
valid near the upper end of the spectrum, i.e., $Cn\ll1$.

\begin{figure*}[t]
\centering
\includegraphics[width=\textwidth]{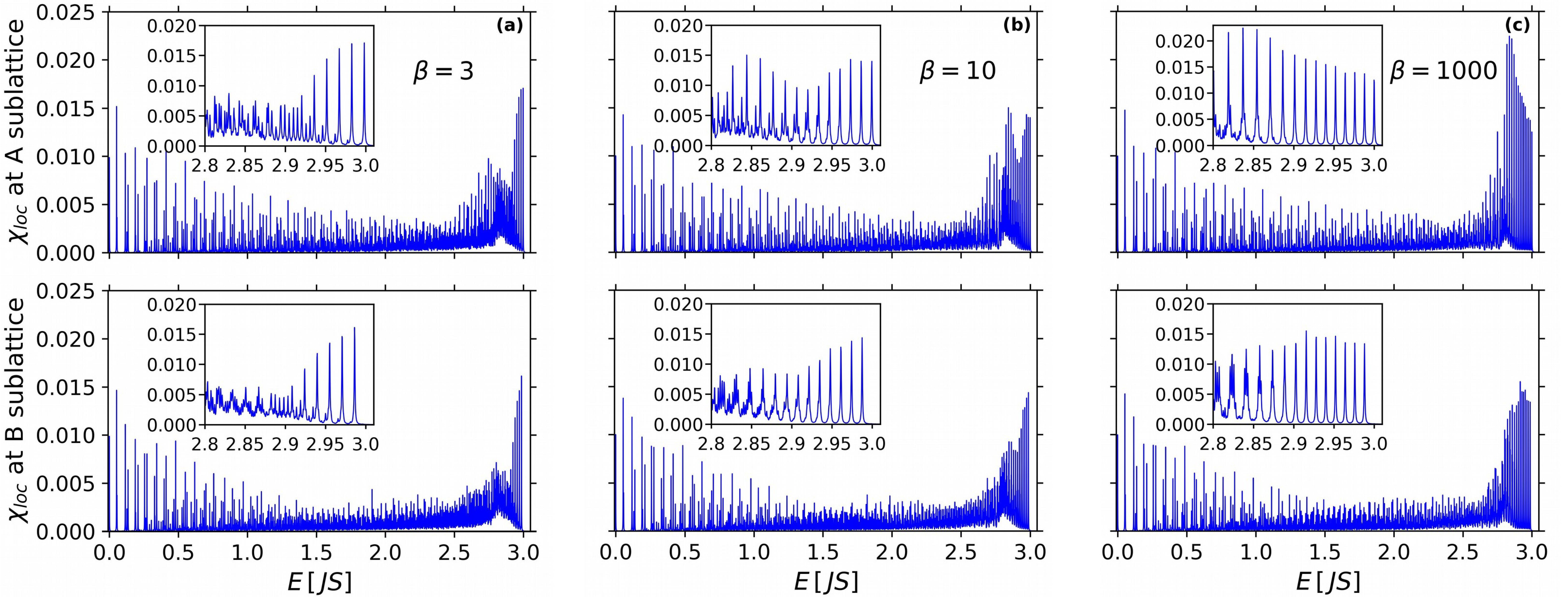}
\caption{Simulation results: Local susceptibility near the sample center on the A (top) and B (bottom) sublattices, obtained for $N=100$, $C/\Cmax=0.4$ and different values of $\beta$.
A Lorentzian broadening of $\eta=10^{-4} JS$ has been applied.
The insets show the upper end of the spectrum, with clearly visible pseudo-Landau levels.
}
\label{fig:ldos1}
\end{figure*}

Eq.~\eqref{magen} is our key result: The $E_n$ represent the energies of highly degenerate magnon pseudo-Landau levels. In contrast to strained graphene, the Landau-level energy decreases with increasing Landau-level index $n$ and the levels are equally spaced -- the latter is related to the unstrained spectrum being quadratic (as opposed to linear) in momentum.



\textit{Emergent supersymmetry.}
Given that the solutions for $n=1, 2, \ldots$ occur in pairs, the degeneracy of the $n>0$ magnon Landau levels is twice as large as that of the $n=0$ level. This implies an emergent supersymmetry, i.e., the single-particle states can be mapped to that of a supersymmetric harmonic oscillator.

We can make this apparent by noting that the single-particle states \eqref{solu}, with energies \eqref{magen}, are the eigenstates of the following Hamiltonian
\begin{equation}
\label{susyosc}
h = 3JS \left[1 - C (F^\dagger F + B^\dagger B)\right]
\end{equation}
where we have introduced the operators
\begin{align}
F &= \sum_m \sum_{n=0}^\infty  |\Phi^+_{n,m}\rangle \langle\Phi^-_{n+1,m}|\,,\\
B &= \sum_m \left( \sum_{n=0}^\infty \!\sqrt{n} |\Phi^+_{n,m}\rangle \langle\Phi^+_{n+1,m}|
   + \sum_{n=1}^\infty \!\sqrt{n} |\Phi^-_{n,m}\rangle \langle\Phi^-_{n+1,m}| \right)
\end{align}
which act in the single-particle Hilbert space, see Fig.~S1 \cite{suppl}. The $|\Phi_{0,m}^+\rangle$ and $|\Phi_{n,m}^\pm\rangle$ are (normalized) states corresponding to the real-space representation \eqref{solu}, with their scalar product defined via a $\Sigma$ norm \cite{suppl}. These states form a complete single-particle basis, and the operators $F$ and $B$ obey the canonical commutation relations for fermions and bosons, respectively, $[F,F^\dagger]_+ = 1$ and $[B,B^\dagger]_- = 1$ \cite{suppl}.

Since both the boson ($B$) and the fermion ($F$) have the same excitation energy, Eq.~\eqref{susyosc} resembles an inverted supersymmetric harmonic oscillator.
We emphasize that the supersymmetry arises from the two-sublattice structure combined with the non-conservation of the bosonic particle number: While the fermionic two-sublattice case yields a spectrum which is particle-hole-symmetric, the bosonic case requires non-negative excitation energies, then implying pairs of degenerate solutions for $n>0$.


\textit{Numerical results.}
To verify our field-theoretic calculation, we now turn to a numerical analysis of the magnon Hamiltonian on finite strained systems. In principle, arbitrarily shaped systems with open boundary conditions can be considered; guided by the insights of Ref.~\onlinecite{rachel16b} we choose triangular-shaped systems with linear size $N$ and $N^2$ sites, as shown in Fig.~\ref{fig:setup}(a).
We employ exchange couplings according to Eqs.~\eqref{displace} and \eqref{ourjij}, yielding a spin-wave Hamiltonian as in Eq.~\eqref{hsw}. We solve the bosonic Bogoliubov problem, amounting to the diagonalization of a non-Hermitian matrix of size $N^2 \times N^2$, using the algorithm outlined in the appendix of Ref.~\onlinecite{wessel05}.

For given system size $N$, the strain $C=\bar C\beta a_0$ has a maximum value $\Cmax$ beyond which some of the exchange couplings become negative due to the linearization in \eqref{ourjij}. Hence, we parameterize the strain by its strength $C/\Cmax$ and consider different values of the magneto-elastic coupling $\beta$. In general, the pseudo-magnetic field will be spatially inhomogeneous, with variations near the edges, and clear Landau-level signatures are expected in the bulk of the sample \cite{peeters13}.

Results for the local magnetic susceptibility, equivalent to the local magnon density of states (DOS), measured near the center of the sample, are shown in Fig.~\ref{fig:ldos1}. Equally spaced Landau levels are clearly visible at the upper end of the spectrum in perfect agreement with Eq.~\eqref{magen}, with the $n=0$ level being present only on the A sublattice \cite{suppl}. The Landau-level spacing scales linearly with applied strain \cite{suppl} as anticipated. Note that the states at the lower spectrum end display small degeneracies only, see also Fig.~\ref{fig:dos1} below, and do not correspond to Landau levels \cite{suppl}, except for the special case in  Fig.~\ref{fig:dos1}(c).

\begin{figure*}[bth]
\centering
\includegraphics[width=\textwidth]{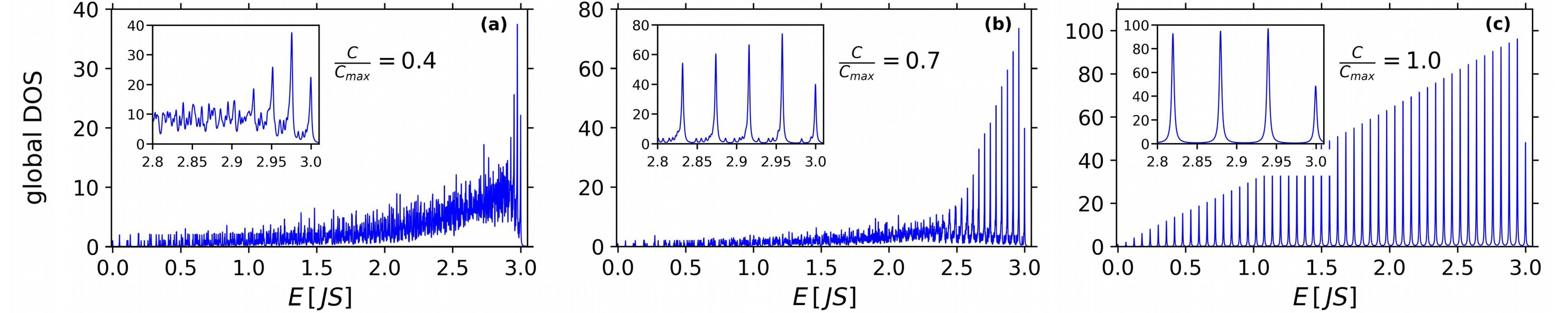}
\caption{Simulation results: Global magnon density of states, obtained for $N=50$, $\beta=1000$ and different values of $C/\Cmax$. A Lorentzian broadening of $\eta=10^{-3} JS$ has been applied ($\eta=1.5 \times 10^{-3} JS$ in panel c).
The insets show the upper end of the spectrum, with the Landau-level spacing scaling linearly with $C/\Cmax$.
}
\label{fig:dos1}
\end{figure*}

As observed in Ref.~\onlinecite{schomerus14} and discussed in detail in Ref.~\onlinecite{rachel16b}, the combined limit of large $\beta$ and small $\bar C$ reduces non-linearities in the pattern of bond strengths, such that pseudo-Landau levels can exist over a large range of energies. For this (artificial) $\beta\to\infty$ limit we show results for the global magnon DOS in Fig.~\ref{fig:dos1}.
Most remarkable is the result for $C=\Cmax$, Fig.~\ref{fig:dos1}(c), where we find perfectly degenerate and equally spaced magnon levels over the full bandwidth. For a system of linear size $N$ there are $N+1$ magnon energies, $E_n=3JS(1-n/N)$, with degeneracies $d_0=N-1$, $d_N=1$, and $d_n=2(N-n)$ for $n=1,\ldots,N-1$. For large $N$ and small $n$, these degeneracies agree with the field-theoretic result derived above and reflect the supersymmetric nature of the spectrum.
For $C<\Cmax$, Fig.~\ref{fig:dos1}(a,b), the energy range of clearly visible magnon Landau levels decreases. Moreover additional peaks appear in between the Landau levels which correspond to magnon states localized near the sample edges, i.e., reflect finite-size effects \cite{suppl}.


\textit{Discussion.}
Having established the existence of magnon Landau levels in strained honeycomb antiferromagnets, we turn to broader aspects.
First, our calculation is controlled in the semiclassical limit of large spin $S$.
Quantum (i.e. $1/S$) corrections lead to higher-order magnon terms in the Hamiltonian. In the present case with collinear order, cubic vertices are forbidden. Self-energy effects from quartic magnon interactions will shift all excitation modes at the upper end of the spectrum in a similar fashion, such that the Landau-level structure remains intact. Broadening effects arising from quartic vertices are typically small. While a full calculation is beyond the scope of this paper, we expect our results to be robust down to $S=1/2$.
Second, the Landau quantization implies that the semiclassical trajectory of a magnon wavepacket centered at $\vec K$ or $\vec K'$ is a circle \cite{unpub}.

The experimental realization of Fig.~\ref{fig:setup}(a) requires to prepare a strained quasi-2D antiferromagnet and to measure its dynamic spin susceptibility. Although challenging, we believe this can be achieved with present-day technology: van-der-Waals-bonded layered magnets such as $\alpha$-RuCl$_3$, CrCl$_3$, or CrI$_3$ can be prepared as ultrathin (even monolayer) samples. Those can be placed on a curved substrate surface to realize suitable strain, akin to what has been done for graphene \cite{levy10}. The excitation spectrum of such a sample may be probed using inelastic Raman scattering or electron-spin-resonance techniques. Realistic values of $\beta=3\ldots 5$ and $\bar C \lesssim 0.2/(Na_0)$ imply a Landau-level spacing of $10^{-3}J$ for a system of $500^2$ atoms.
Alternatively, a strained antiferromagnet may be realized in a cold-atom quantum simulator. This requires to generate a Hubbard model in the Mott regime \cite{greif16} on a suitable inhomogeneous optical lattice. More generally, all quantum simulators for fermionic Hubbard models could in principle be used \cite{hensgens17}, in particular when an arbitrary two-dimensional lattice can be simulated \cite{salfi16}.


\textit{Summary.}
We have demonstrated how to engineer a novel state of matter, where magnon excitations of a collinearly ordered antiferromagnet display highly degenerate Landau levels at the top of the spectrum. This realization of strain-induced Landau levels fundamentally differs from previous ones, as the starting spectrum displays a quadratic (instead of linear) dispersion. Consequently, the spectrum features equidistant Landau levels, and moreover realizes emergent supersymmetry.
We have also discovered that the combined limit of strong magneto-elastic coupling and weak lattice deformations leads to perfectly quantized magnon Landau levels over the full range of magnon energies.

Generalizations to other lattice structures will be discussed elsewhere. We anticipate that adapting the scheme developed in Ref.~\onlinecite{rachel16b} will enable us to engineer magnon Landau levels also in collinear diamond-lattice antiferromagnets. Strain engineering of non-collinear antiferromagnets \cite{attig17} will not only modify the excitation spectrum, but also lead to strain-dependent reference states, possibly generating entirely new types of order -- this exciting phenomenology is left for future work.


We thank D. Arovas, J. Attig, L. Fritz, I. Goethel, A. Rosch, and S. Trebst for discussions as well as collaborations on related work.
We acknowledge financial support from the DFG through SFB 1143 and the W\"urzburg-Dresden Cluster of Excellence on Complexity and Topology in Quantum Matter -- \textit{ct.qmat} (EXC 2147, project-id 39085490) as well as by the IMPRS on Chemistry and Physics of Quantum Materials. SR acknowledges an ARC Future Fellowship (FT180100211).



\begin{thebibliography}{}
\bibitem{classnote}
In the semiclassical limit, $S\to\infty$, the Hamiltonian \eqref{hk} on the honeycomb lattice has a N\'eel ground state for any choice of $J_{ij}>0$ which follows from energy minimization.
\end{thebibliography}
\end{document}


\title{
Supplemental material for \\
Magnon Landau levels and emergent supersymmetry in strained antiferromagnets
}

\author{Mary Madelynn Nayga}
\affiliation{Institut f\"ur Theoretische Physik and W\"urzburg-Dresden Cluster of Excellence ct.qmat, Technische Universit\"at Dresden,
01062 Dresden, Germany}
\affiliation{Max-Planck-Institut f\"ur Chemische Physik fester Stoffe, N\"othnitzer Str. 40,
01187 Dresden, Germany}
\author{Stephan Rachel}
\affiliation{School of Physics, University of Melbourne, Parkville, VIC 3010, Australia}
\author{Matthias Vojta}
\affiliation{Institut f\"ur Theoretische Physik and W\"urzburg-Dresden Cluster of Excellence ct.qmat, Technische Universit\"at Dresden,
01062 Dresden, Germany}

\date{\today}

\maketitle


\section{Continuum-limit spin-wave theory}
\label{sec:continuum}

In this section we derive the continuum-limit Hamiltonian for spin waves near momenta $\vec K$, $\vec K'$ of a weakly strained honeycomb-lattice Heisenberg antiferromagnet, Eq.~\eqhstrain of the main text.

\subsection{Heisenberg antiferromagnet without strain}
\label{subsec:nostrain}

To set the stage, we start with the unstrained case. Introducing Holstein-Primakoff bosons describing fluctuations above the collinear N\'eel state,
$S^z_{i,A} = S-a_i^\dagger a_i$,
$S^z_{j,B} = -S+b_j^\dagger b_j$,
$S^-_{i,A} = a_i^\dagger \sqrt{2S - a_i^\dagger a_i}$,
$S^+_{j,B} = b_j^\dagger \sqrt{2S - b_j^\dagger b_j}$
on $A$ and $B$ sublattices, respectively,
the lattice spin-wave Hamiltonian on a bipartite lattice is
\begin{equation}
\label{hsw}
\HSW = S J \sum_{\langle ij\rangle} \left(a_i^\dagger a_i + b_j^\dagger b_j + a_i b_j + a_i^\dagger b_j^\dagger \right)
\end{equation}
as given in Eq.~\eqhsw in the main text. The nearest-neighbor vectors $\vec{\delta}_{1,2,3} = \vec{r}_j - \vec{r}_i$ on the honeycomb lattice are
\begin{equation}
\vec{\delta_1} = a_0\big(\dfrac{\sqrt{3}}{2},\dfrac{1}{2}\big),~
\vec{\delta_2} = a_0\big(-\dfrac{\sqrt{3}}{2},\dfrac{1}{2}\big),~
\vec{\delta_3} = a_0(0,-1)
\end{equation}
where $a_0$ is the lattice constant.
Focussing on magnon modes with momenta in the vicinity of
$\vec K= 2\pi/(3a_0)(1/\sqrt{3},1)$,
we can pass to the continuum limit by expanding the bosonic lattice operators $a_i$, $b_j$ near $\vec K$ according to
\begin{eqnarray}
a(\vec{r}) &\approx & e^{i \vec{K}  \cdot \vec{r}} \psi_a(\vec{r})\,, \nonumber \\
b(\vec{r}) &\approx & e^{-i \vec{K} \cdot \vec{r}} \psi_b(\vec{r})~,
\label{opexpand}
\end{eqnarray}
where $\psi_{a,b}(\vec{r})$ are slowly-varying bosonic fields, each with dimension of [length]$^{-1}$. Note that the Bogoliubov nature of the problem dictates a coupling between $a_{\vec K}$ and $b^\dagger_{-\vec K}$, which necessitates the different momentum signs in Eq.~\eqref{opexpand}.
It is convenient to introduce the two-vector $\Psi=(\psi_a(\vec r),\psi_b^\dagger(\vec r))^T$.
A gradient expansion of the non-local terms yields a Hamiltonian of the form
\begin{eqnarray}
\label{hcontgen}
\HSW &=&  \int d^2r \, \, \Psi^\dagger (\vec{r}) \, \mathcal{H}(\vec r) \, \Psi (\vec{r})\,.
\end{eqnarray}
In the absence of strain, the Hamiltonian density $\mathcal{H}(\vec r)$ reads
\begin{equation}
\mathcal{H}_0(\vec r) = SJ \sum_{j=1}^3
\begin{pmatrix}
1 & e^{i\vec{K}\cdot \vec{\delta_j}}\left(1\!+\!\vec{\delta_j} \cdot\vec{\nabla} \right) \\
e^{-i\vec{K}\cdot \vec{\delta_j}}\left(1\!-\!\vec{\delta_j} \cdot\vec{\nabla} \right) & 1
\end{pmatrix}
.
\end{equation}
Using the explicit forms of $\vec K$ and $\vec\delta_j$ we obtain
\begin{equation}
\label{hdens0}
\mathcal{H}_0(\vec r)= 3JS
\begin{pmatrix}
 1 & -\frac{e^{i\pi /3}}{2} a_0 \left( p_x - i p_y \right) \\
- \frac{e^{-i\pi /3}}{2} a_0 \left(p_x + i p_y \right)  & 1
\end{pmatrix}
\end{equation}
with $p_x=-i\partial_x$, $p_y=-i\partial_y$.
This Hamiltonian can be diagonalized by subsequent Fourier and Bogoliubov transformations, leading to
\begin{equation}
\w_k = 3JS \sqrt{1 - k^2 a_0^2/4}~ \approx 3JS( 1 - k^2 a_0^2/ 8)
\end{equation}
which coincides with the small-momentum expansion of the lattice magnon dispersion, Eq.~\eqspecsw of the main text, where $\vec k$ is now the momentum relative to $\vec K$.

\subsection{Heisenberg antiferromagnet on strained honeycomb lattice}
\label{subsec:with strain}

We now derive the continuum Hamiltonian for arbitrary forms of (small) lattice deformations. The deformed lattice, with coordinates $\vec r'$, is related to the undeformed lattice as follows:
\begin{eqnarray}
\vec{r} \quad &\rightarrow &  \quad \vec{r'} = \vec{r} + \vec{\dis}(\vec{r}) \\
\label{delta'}
\vec{r} + \vec{\delta_j} \quad  &\rightarrow & \quad \vec{r'} + \vec{\delta_j'} = (\vec{r} +\vec{\delta_j}) + \vec{\dis}(\vec{r} + \vec{\delta_j}),
\end{eqnarray}
where $\vec{\dis}(\vec{r})$ is a slowly-varying displacement field. The new lattice vectors are
\begin{equation}
\label{eqn: new delta}
\vec{\delta_j'} \approx \vec{\delta_j} + (\vec{\delta_j} \cdot \vec{\nabla}_{r'})\vec{\dis}(\vec{r'})
\end{equation}
and the corresponding exchange interaction $J(\vec{\delta_j'})$ reads
\begin{align}
\label{eqn: new J}
J(\vec{\delta_j'})
&\approx J(\vec{\delta_j} + (\vec{\delta_j} \cdot \vec{\nabla}_{r'})\vec{\dis}) \notag\\
&\approx J(\vec{\delta_j}) + (\vec{\delta_j} \cdot \vec{\nabla}_{r'})\vec{\dis} \cdot \vec{\nabla}_{r'} J(\vec{\delta_j})
\end{align}
The last term describes the dependence of $J(\vec\delta)$ on the bond length $|\vec \delta|$ which we parameterize by
$\beta = - d\ln J / d\ln |\delta|$, to be taken at $|\delta|=a_0$ -- this coincides with the definition of $\beta$ in Eq.~\eqourjij of the main text. In general, $\beta$ is a materials parameter of order unity, for graphene \cite{neto09} $\beta=3.3$.

Inserting all expressions into the magnon Hamiltonian, we can expand in $\dis$ and field gradients. To lowest non-trivial order, the displacement enters via the strain tensor $\uu_{ij} = (\partial_i \dis_j + \partial_j \dis_i)/2$.
Keeping terms up to first order in $\uu$ and neglecting products of $\uu$ and field gradients, we arrive (after renaming $\vec r'\to\vec r$) at a continuum Hamiltonian of the form \eqref{hcontgen}, with a Hamiltonian density
\begin{equation}
\mathcal{H}(\vec r)=\mathcal{H}_0(\vec r)+\mathcal{H}_1(\vec r)
\end{equation}
where $\mathcal{H}_0(\vec r)$ corresponds to the unstrained system and is in Eq.~\eqref{hdens0}, and $\mathcal{H}_1(\vec r)$ is linear in strain and given by
\begin{widetext}
\begin{equation}
\label{hdens1}
\mathcal{H}_1(\vec r) = SJ \sum_{j=1}^3
\begin{pmatrix}
-\dfrac{\beta}{a_0^2}\vec{\delta_j}\cdot \uu \cdot \vec{\delta_j}  &  e^{i\vec{K} \cdot\vec{\delta_j}} \left( -\dfrac{\beta}{a_0^2}\vec{\delta_j}\cdot \uu \cdot \vec{\delta_j} + i \vec{K}\cdot \uu \cdot \vec{\delta_j} \right) \\
e^{-i\vec{K} \cdot \vec{\delta_j}} \left(-\dfrac{\beta}{a_0^2}\vec{\delta_j}\cdot \uu \cdot \vec{\delta_j} - i \vec{K}\cdot \uu \cdot \vec{\delta_j} \right) & -\dfrac{\beta}{a_0^2}\vec{\delta_j}\cdot \uu \cdot \vec{\delta_j}
\end{pmatrix}
\end{equation}
Again, using the explicit expressions for $\vec K$ and $\vec\delta_j$ and performing a gauge transformation $\psi_a \to \psi_a e^{-i\pi /3}$, we can bring the full Hamiltonian density into the form
\begin{equation}
\label{hstrain}
\mathcal{H}(\vec r) = 3JS
\begin{pmatrix}
1+a & -\frac{a_0}{2} (p_x + A_x - i p_y - i A_y) \\
-\frac{a_0}{2} (p_x + A_x + i p_y + i A_y) & 1+a
\end{pmatrix}
\end{equation}
\end{widetext}
which describes honeycomb-lattice magnons near wavevector $\vec K$ and is identical to Eq.~\eqhstrain of the main text. The strain-induced contributions are
\begin{equation}
\label{pseudoa}
a = - \frac{\beta}{2} (\uu_{xx}+\uu_{yy})
\end{equation}
and
\begin{equation}
\label{pseudoA}
\vec{A} = \dfrac{\beta}{2a_0}
\begin{pmatrix}
\uu_{xx}-\uu_{yy}  \\
-2\uu_{xy}
\end{pmatrix} \,
+ \dfrac{2\sqrt{3}\pi}{9a_0}
\begin{pmatrix}
\uu_{xx}+\sqrt{3}\uu_{xy}  \\
\sqrt{3}\uu_{yy}+\uu_{xy}
\end{pmatrix}.
\end{equation}
The diagonal correction $a$ \eqref{pseudoa} renormalizes the magnon on-site energy and thus can be interpreted as bandwidth renormalization from strain, e.g., it is non-zero for a homogeneous expansion or compression of the lattice.
The pseudo-vector potential $\vec A$ \eqref{pseudoA} is identical to that obtained for strained graphene which has been discussed extensively in numerous papers.\cite{pereira09,kitt12,masir13} We note that often  only the $\beta$-dependent (i.e. first) piece of $\vec{A}$ is reported; it is this term which is responsible for generating a homogeneous pseudo-magnetic field from triaxial strain. The second term, denoted $A_{\vec K}$ in the main text, is independent of the actual change of coupling constants and thus represents a lattice correction reflecting the distortion of the Brillouin zone, see Refs.~\onlinecite{kitt12,masir13} for a discussion.


\section{Continuum theory for magnon pseudo-Landau levels}
\label{sec:triaxial strain}

\subsection{Magnon pseudo-Landau levels from triaxial strain}

For the Dirac electron of graphene, it has been shown that a homogeneous pseudo-magnetic field can be generated by triaxial strain,\cite{Fogler08,Guinea10,Vozmediano10} with the displacement field
\begin{equation}
\label{triaxial strain}
\vec{\dis}(\vec{r}) = \bar{C} \left(2xy, x^2 - y^2 \right)^T,
\end{equation}
where the constant $\bar{C}$ parameterizes the lattice distortion. In fact, the $\beta$-dependent piece of the pseudo-vector potential \eqref{pseudoA} yields a constant pseudo-magnetic field $B\propto \beta\bar C$, and for realistic values of $\beta$ ($\beta=3.3$ for graphene) the $\beta$-independent piece produces small corrections only.\cite{kitt12,masir13}
To keep the analytical calculation tractable, we therefore keep the $\beta$-dependent piece of $\vec A$ only. This is formally justified in the combined limit of $\beta\to\infty$, $\bar C\to 0$,\cite{schomerus14,rachel16b} and our independent numerical results confirm that the $\beta$-independent lattice corrections do not qualitatively change the conclusions even for $\beta=3$.

Combining Eqs.~\eqref{hstrain} and \eqref{triaxial strain}, the Hamiltonian density for the triaxially strained honeycomb antiferromagnet can thus be written as
\begin{equation}
\mathcal{H}(\vec r) = 3JS
\begin{pmatrix}
 1 & - \frac{a_0}{2}\left(\Pi_x - i \Pi_y \right) \\
-\frac{a_0}{2} \left(\Pi_x+i\Pi_y\right)  & 1
\end{pmatrix}
\end{equation}
where $\vec\Pi = \vec p + \vec A$ reads explicitly
\begin{equation}
\Pi_x = p_x +\frac{2\beta \bar{C}}{a_0}y,~~\Pi_y = p_y - \frac{2\beta \bar{C}}{a_0}x,
\end{equation}
with the commutation relation $[\Pi_x , \Pi_y]_- = 4i \beta \bar{C}/a_0$.

At this point, we combine the strain parameters $\beta$ and $\bar C$ into the dimensionless strain measure $C = \bar{C}\beta a_0$. Following the standard treatment of Landau-level problems in symmetric gauge, it is convenient to introduce two types of bosonic ladder operators
\begin{align}
\alpha &= \left[p_x + y \frac{2C}{a_0^2} + i (p_y - x \frac{2C}{a_0^2}) \right] \frac{a_0}{\sqrt{8C}} \,,\\
\gamma &= \left[p_x - y \frac{2C}{a_0^2} - i (p_y + x \frac{2C}{a_0^2}) \right] \frac{a_0}{\sqrt{8C}},
\end{align}
and we assumed $C>0$. The Hamiltonian density becomes
\begin{equation}
\label{hstrain2}
\mathcal{H}(\vec r) = 3JS
\begin{pmatrix}
1 & -{\sqrt{2C}} \alpha^\dagger  \\
-{\sqrt{2C}} \alpha & 1
\end{pmatrix}
.
\end{equation}
The Hamiltonian is diagonalized by a continuum version of the bosonic Bogoliubov transformation.\cite{wessel05} This amounts to finding the eigenfunctions of the non-Hermitian matrix $\Sigma \mathcal{H}(r)$ where $\Sigma = {\rm diag}(1,-1)$. This eigenvalue problem reads
\begin{equation}
\label{eq: eigenvalue equation at K}
\begin{pmatrix}
1 & -\sqrt{2C} \alpha^\dagger \\
\sqrt{2C} \alpha & - 1
\end{pmatrix}
\begin{pmatrix}
\phi_a(r) \\
\phi_b(r)
\end{pmatrix}
= \lambda
\begin{pmatrix}
\phi_a(r) \\
\phi_b(r)
\end{pmatrix}
\end{equation}
where $\lambda$ corresponds to a single-particle energy in units of $3JS$.
The two differential equations
\begin{align}
\sqrt{2C} \alpha^\dagger \phi_b &= (1-\lambda) \phi_a, \\
\sqrt{2C} \alpha \phi_a &= (1+\lambda) \phi_b
\end{align}
can be decoupled to yield
\begin{eqnarray}
\alpha^\dagger \alpha \phi_a &=& \frac{1}{2 C} \left( 1 - \lambda^2 \right) \phi_a\,.
\end{eqnarray}
Using $\alpha^\dagger \alpha = n$ we have $\lambda_n^\pm = \pm \sqrt{1-2Cn}$, and the resulting energy quantization is
\begin{equation}
\label{magen}
E_n = 3JS \sqrt{1 - 2 C n} \approx 3JS (1 - C n),~~n=0,1,2,\ldots
\end{equation}
reflecting equally spaced pseudo-Landau levels at the upper end of the spectrum, as given in the main text. Small strain corresponds to $C\ll 1$, the condition of small deviations from the upper end of spectrum implies in addition $Cn\ll 1$.
The wavefunctions corresponding to energy $E_n$ are (up to a global normalization factor)
\begin{align}
\phi_{a} &= c_n^\pm \phi_n \equiv \frac{\sqrt{2Cn}}{1-\lambda_n^\pm} \phi_n \\
\phi_{b} &= \phi_{n-1}
\end{align}
where $\phi_n = (n!)^{-1/2} (\alpha^\dagger)^n \phi_0$ with $\alpha\phi_0=0$. The second pair of ladder operators $\gamma^\dagger$, $\gamma$ induces a degeneracy of all levels corresponding to $m=\gamma^\dagger\gamma$. Combining $\phi_a$ and $\phi_b$ into a spinor $\Phi=(\phi_a,\phi_b)$ yields the single-particle wavefunctions given in Eq.~~\eqsolu of the main text (where $Cn\ll 1$ has been used). In particular, the topmost ($n=0$) Landau level has non-zero amplitude on the A sublattice only. Also, $Cn\ll 1$ implies $c_n^+ \gg 1$ and $c_n^- \ll 1$, i.e., the two sets of solutions for the $n>0$ Landau levels differ in their weight distribution between A and B sublattices.

Performing the expansion about
$\vec K'= -2\pi/(3a_0)(1/\sqrt{3},1)$
instead of $\vec K$ changes the sign of the pseudo-vector potential and also interchanges the role of $\alpha$ and $\gamma$ boson operators. The energy spectrum remains unchanged, as is the property that the $n=0$ Landau level lives on the A sublattice only. The explicit breaking of sublattice symmetry arises from the strain pattern: Inverting the sign of the triaxial strain, $C\leftrightarrow -C$, interchanges the role of A and B sublattices.

\begin{figure*}[htb]
\centering
\includegraphics[width=0.7\textwidth]{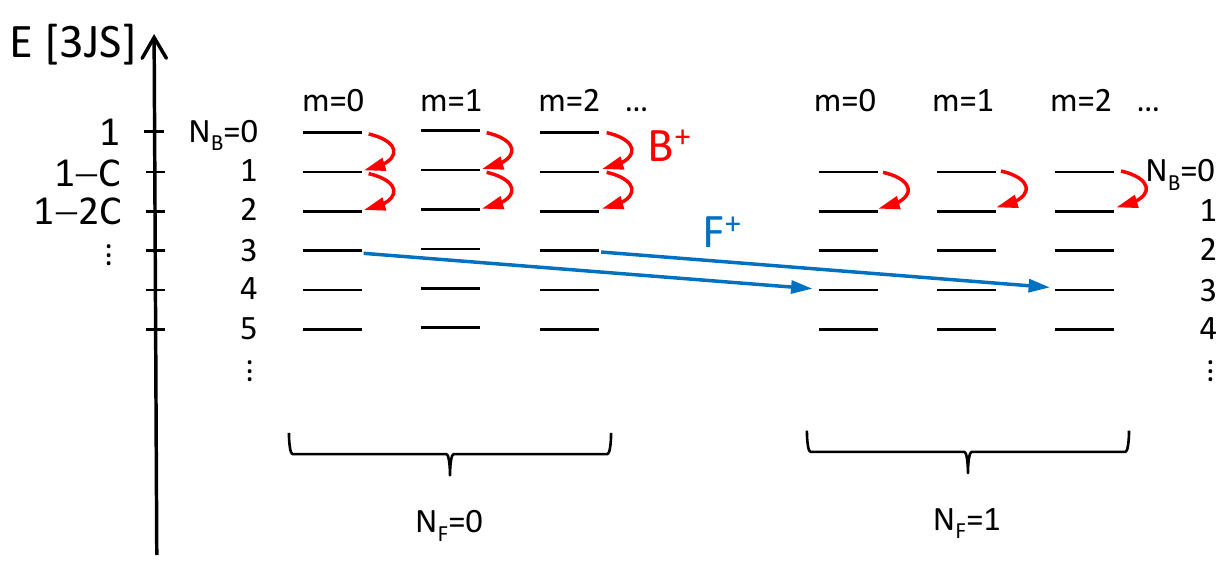}
\caption{
Single-magnon level scheme, illustrating its supersymmetric character.
The colored arrows indicate the action of the $B^\dagger$ and $F^\dagger$ operators; the corresponding occupation numbers are $N_B=B^\dagger B$ and $N_F=F^\dagger F$.
}
\label{fig:susy}
\end{figure*}

\subsection{Emergent supersymmetry}

The single-particle spectrum \eqref{magen}, together with the pairwise degeneracy of the $n>0$ levels, can be brought in supersymmetric form. It is identical to the eigenvalue spectrum of the Hamiltonian $h = 3JS[1-C (F^\dagger F + B^\dagger B)]$, i.e., an inverted supersymmetric harmonic oscillator, Fig.~\ref{fig:susy}. The corresponding operators $F$ and $B$ act in the single-particle (i.e. single-magnon) space and are defined as
\begin{align}
F &= \sum_m \sum_{n=0}^\infty  |\Phi^+_{n,m}\rangle \langle\Phi^-_{n+1,m}|\,,\\
B &= \sum_m \left( \sum_{n=0}^\infty \sqrt{n} |\Phi^+_{n,m}\rangle \langle\Phi^+_{n+1,m}|
   + \sum_{n=1}^\infty \sqrt{n} |\Phi^-_{n,m}\rangle \langle\Phi^-_{n+1,m}| \right).
\end{align}
As these states derive from a bosonic Bogoliubov problem, their scalar product involves $\Sigma = {\rm diag}(1,-1)$, i.e., for two states with the real-space representation $\langle \vec r|\Phi\rangle = (u(\vec r), v(\vec r))^T$, $\langle \vec r|\Phi'\rangle = (u'(\vec r), v'(\vec r))^T$ the scalar product is defined as
\begin{equation}
\langle \Phi|\Phi'\rangle = \int d^2 r (u^\ast u' - v^\ast v')
\end{equation}
and $\langle \Phi|\Phi\rangle = \int d^2 r (|u|^2 - |v|^2)$ is the $\Sigma$ norm.
The eigenstates, with proper normalization, obey $\langle \Psi^+_{n,m}|\Psi^+_{n',m'}\rangle =\delta_{nn'}\delta_{mm'}$, $\langle \Psi^-_{n,m}|\Psi^-_{n',m'}\rangle =\delta_{nn'}\delta_{mm'}$, $\langle \Psi^+_{n,m}|\Psi^-_{n',m'}\rangle = 0$. As a result
\begin{align}
&[F,F^\dagger]_+ = [B,B^\dagger]_- \\
&=
\sum_m \left(
  \sum_{n=0}^\infty |\Psi^+_{n,m}\rangle \langle\Psi^+_{n,m}|
+ \sum_{n=1}^\infty |\Psi^-_{n,m}\rangle \langle\Psi^-_{n,m}|
\right)\,. \notag
\end{align}
The latter is a unity operator in the single-particle Hilbert space, i.e., $F$ and $B$ are canonical fermion and boson operators, respectively.
Note that all eigenstates of the Hamiltonian $h$ have an additional degeneracy corresponding to $m$: for instance, the vacuum of $B$ and $F$ is spanned by the $\{|\Phi^+_{0,m}\rangle\}$.


\section{Strain and Goldstone modes}

While we have established that magnons near $\vec{K}$ and $\vec{K'}$, i.e., at the upper end of the spectrum, acquire a strain-induced pseudo-vector potential, it is also interesting to think about the fate of the low-energy part of the magnon spectrum under the influence of strain. For the unstrained system, these low-energy (Goldstone) magnons display a linear dispersion centered around $\vec{\Gamma}=(0,0)$, see Fig.~1 and Eq.~(3) of the main text.

The honeycomb-lattice Heisenberg model has a stable antiferromagnetic ground state with spontaneously broken SU(2) symmetry; this statement remains true in the presence of any type of (small) strain. For a finite-size system Goldstone's theorem then dictates that there must be a \emph{single} spin-wave mode at zero energy (while the details of the spectrum at finite energies depend on the strain pattern).
Importantly, this implies that a Landau-level structure of the low-energy magnons is forbidden to exist, because a Landau-level structure would imply either highly degenerate magnons at zero energy (corresponding to an instability of the system which, however, is not present) or to no magnons at zero energy at all (in conflict with Goldstone's theorem).
From this we conclude that Goldstone modes cannot acquire a pseudo-vector potential under strain on general grounds.

For the honeycomb lattice antiferromagnet, we have explicitly performed the relevant continuum-limit calculation, i.e., a calculation similar to that in Sec.~\ref{sec:continuum} above, but expanding about $\vec{\Gamma}$ instead of $\vec{K}$. The resulting Hamiltonian is lengthy and not particularly illuminating and thus will not be reproduced here. However, it is clear that it cannot be cast into a minimal-coupling form required for a pseudo-vector potential. This is consistent with the reasoning above and also consistent with our numerical results for general values of strain. We also conclude that the strain engineering proposed here does not universally work for every ordered antiferromagnet.


\begin{figure*}[t]
\centering
\includegraphics[width=\textwidth]{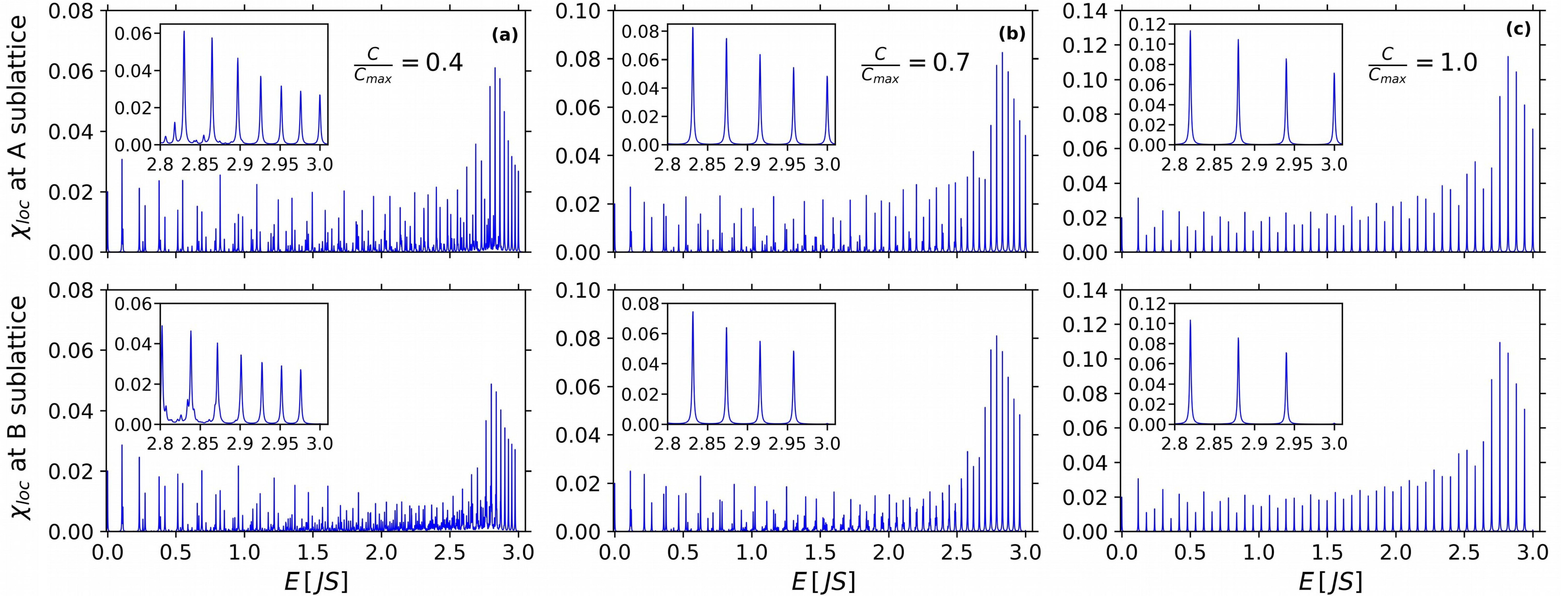}
\caption{Local susceptibility near the sample center on the A (top) and B (bottom) sublattices, obtained for $N=50$, $\beta=1000$ and different values of $C/\Cmax$.
A Lorentzian broadening of $\eta=10^{-3} JS$ has been applied.
The insets show the upper end of the spectrum. The corresponding global DOS is in Fig.~3 of the main paper.
The Landau-level spacing can be seen to be proportional to the strain $C$.
}
\label{fig:ldos1}
\end{figure*}

\section{Numerical simulations}

\subsection{Implementation}

Our numerical calculations follow the procedure outlined in Ref.~\onlinecite{wessel05}, and we summarize the most important steps. The Hamiltonian \eqref{hsw} can be written as
\begin{equation}
\HSW = \bar{a}^\dag M \bar{a}
\end{equation}
with the column vector $\bar{a}$ of size $N=N_A+N_B$ and
\begin{equation}
 \bar{a}^\dag=(a^\dag_1,...,a^\dag_{N_A},b_1,...,b_{N_B}),
\end{equation}
$N_{A,B}$ the number of sites in the A, B sublattices,  and $M$ being the Hermitian $N\times N$ coupling matrix. The Hamiltonian is diagonalized by the Bogoliubov transformation
\begin{equation}
 \label{eq:T}
 \bar{a}=T\bar{\beta},
\end{equation}
with
\begin{equation}
 \bar{\beta}^\dag=(\beta^\dag_1,...,\beta^\dag_{N_+},\beta_{N_++1},...,\beta_{N})
\end{equation}
where $N_+$ and $N_-$ are the numbers of positive and negative eigenvalues of $\Sigma M$, see below, with $N_++N_-=N$. In order to preserve bosonic commutation relations, the transformation
$T$ needs to fulfill
\begin{equation}
 \label{eq:bcr}
 T \Gamma T^\dag = \Sigma,
\end{equation}
with matrices $\Gamma$ and $\Sigma$ defined by
\begin{equation}
\label{eq:SL}
 \Gamma= \begin{pmatrix} 1_{N_+} & 0 \\ 0 & -1_{N_-} \end{pmatrix},
 \quad\Sigma= \begin{pmatrix} 1_{N_A} & 0 \\ 0 & -1_{N_B} \end{pmatrix}.
\end{equation}
$T$ diagonalizes $\mathcal{H}$, i.e.,
\begin{equation}
 T^\dag M T= \Omega, \quad \Omega={\rm diag}(\Omega_1,...,\Omega_{N}),
\end{equation}
and the $\w_k = |\Omega_k|$ are the physical eigenfrequencies of the problem.
In practice, we solve the non-Hermitian eigenvalue problem
\begin{equation}
 \Sigma M T=T\Gamma \Omega
\end{equation}
using the algorithm outlined in the appendix of Ref.~\onlinecite{wessel05}.
The density of states of the spin-wave excitation spectrum (DOS) is simply
\begin{equation}
\rho(\omega)=\sum_k \delta(\omega-\omega_k).
\end{equation}

The transformation matrix $T$ can be decomposed into the $N_A\times N_\pm$ matrices $A_\pm$ and the $N_B \times N_\pm$ matrices $B_\pm$ as
\begin{equation}
 T=\begin{pmatrix} A_+ & A_- \\ B_+ & B_- \end{pmatrix}
\end{equation}
to define $N\times N$ matrices
\begin{equation}
 U=\begin{pmatrix} A_+ & 0 \\ 0 & B_-^* \end{pmatrix}, \quad V=\begin{pmatrix} 0 & A_- \\ B_+^* & 0 \end{pmatrix}.
\end{equation}
With these we can determine the real-space dynamic spin structure factor
\begin{equation}
 S_{ij}^{\perp}(\omega)=\frac{1}{2}\int_{-\infty}^{\infty} dt\: e^{i\omega t} \langle S^+_i(t) S^-_j(0) + S^-_i(t) S^+_j(0)
\rangle.
\end{equation}
In linear spin-wave theory this reads
\begin{equation}
S_{ij}^{\perp}(\omega)=S \sum_k \left( U_{ik} U^*_{jk} + V^*_{ik} V_{jk} \right) \delta(\omega-\omega_k),
\end{equation}
such that the local dynamic spin susceptibility at $T=0$ is given by
\begin{equation}
\label{chiloc}
\chi_{ii}(\w) = S \sum_k (|U_{ik}|^2 + |V_{ik}|^2) \delta(\w - \w_k)\,.
\end{equation}

\subsection{Additional results}

\begin{figure}[htb]
\centering
\includegraphics[width=0.9\columnwidth]{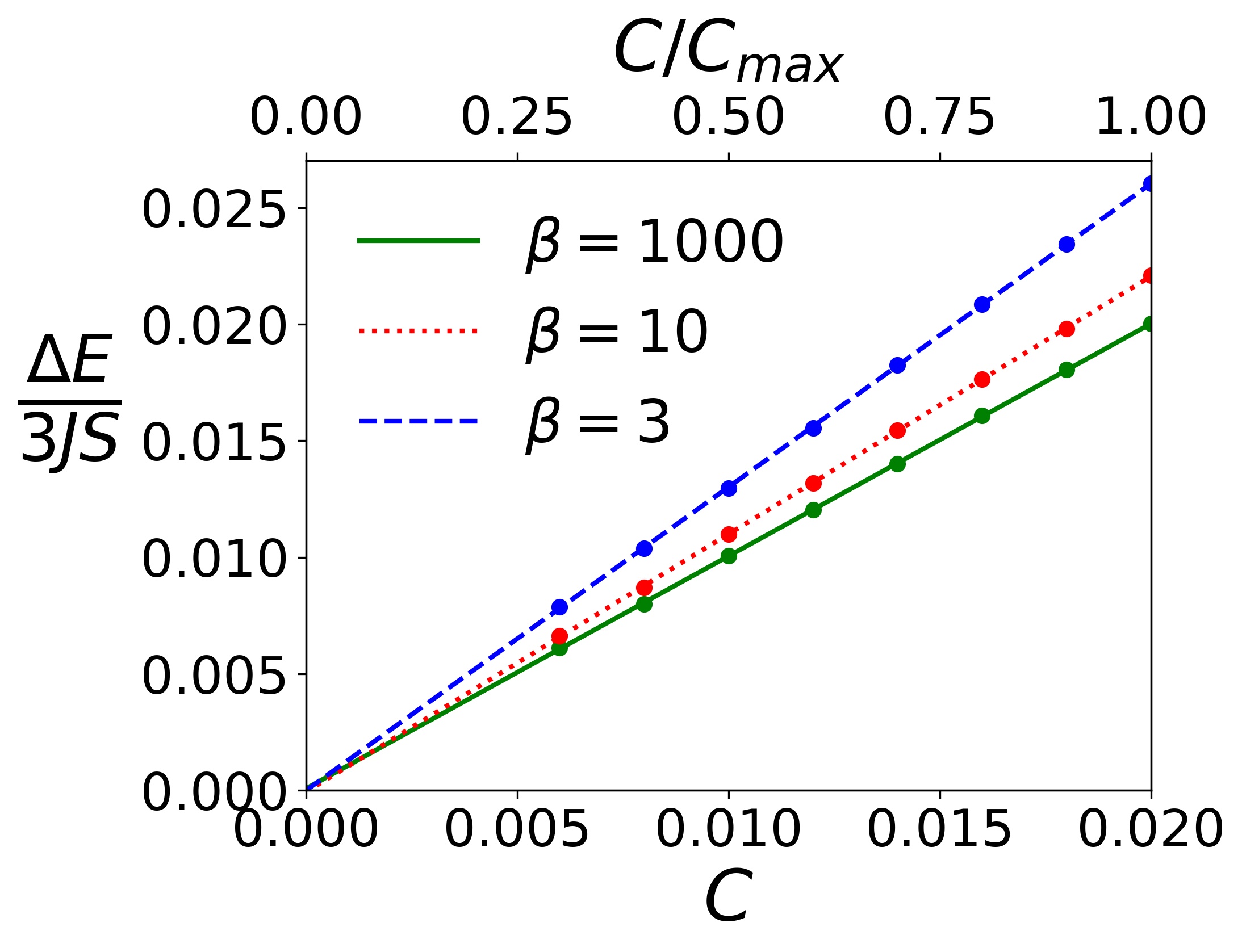}
\caption{Landau-level spacing, $\Delta E$ in units of $3JS$, as function of dimensionless strain $C=\bar C\beta a_0$, calculated as the energetic difference between the two highest Landau levels for $N=50$ (where $\Cmax=0.02$) and different values of $\beta$. The dashed lines are linear fits with slopes $1.301$, $1.106$, and $0.997$ for $\beta=3,10,1000$, respectively. The field-theory prediction, valid in the limit of large $\beta$, is $\Delta E = 3JS C$, see Eq.~\eqref{magen}.
}
\label{fig:spacing}
\end{figure}

\begin{figure*}[t!]
\includegraphics[width=\textwidth]{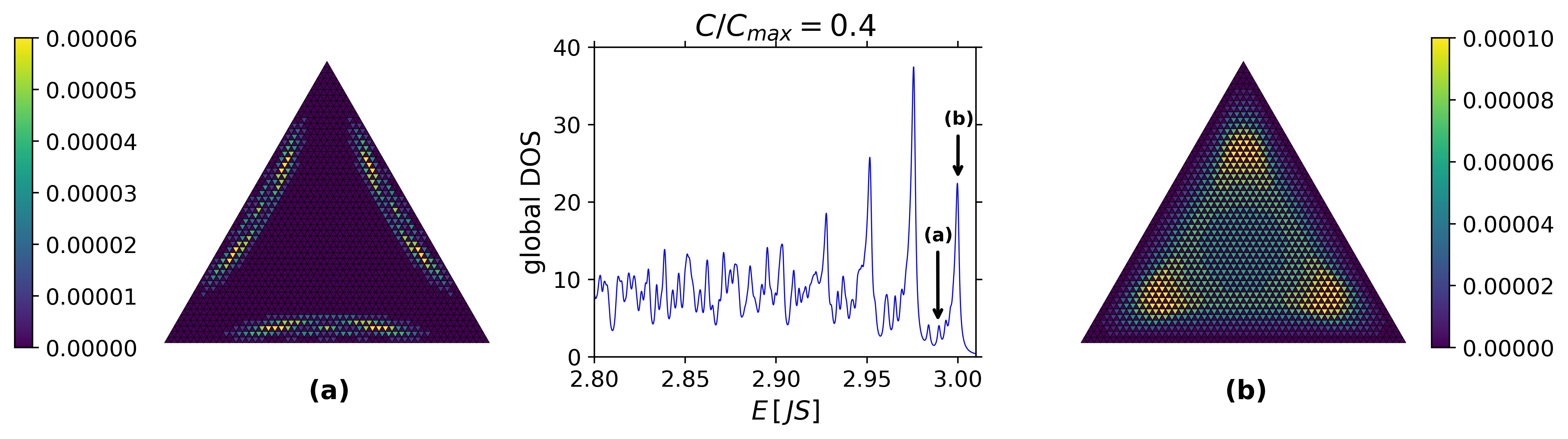}
\caption{
Spatial profile of eigenmodes: Color-coded local dynamic susceptibility $\chi_{ii}(\w)$ for every site $i$ and integrated over a narrow window of energies $\w$ (width $4\times10^{-3}JS$), for two peaks in the global DOS labelled a (left) and b (right); the global DOS is shown in the middle panel.
All data are for $N=50$, $\beta=1000$, and $C/\Cmax=0.4$.
}
\label{fig:ldosprofile1}
\end{figure*}

\begin{figure*}[t!]
\includegraphics[width=\textwidth]{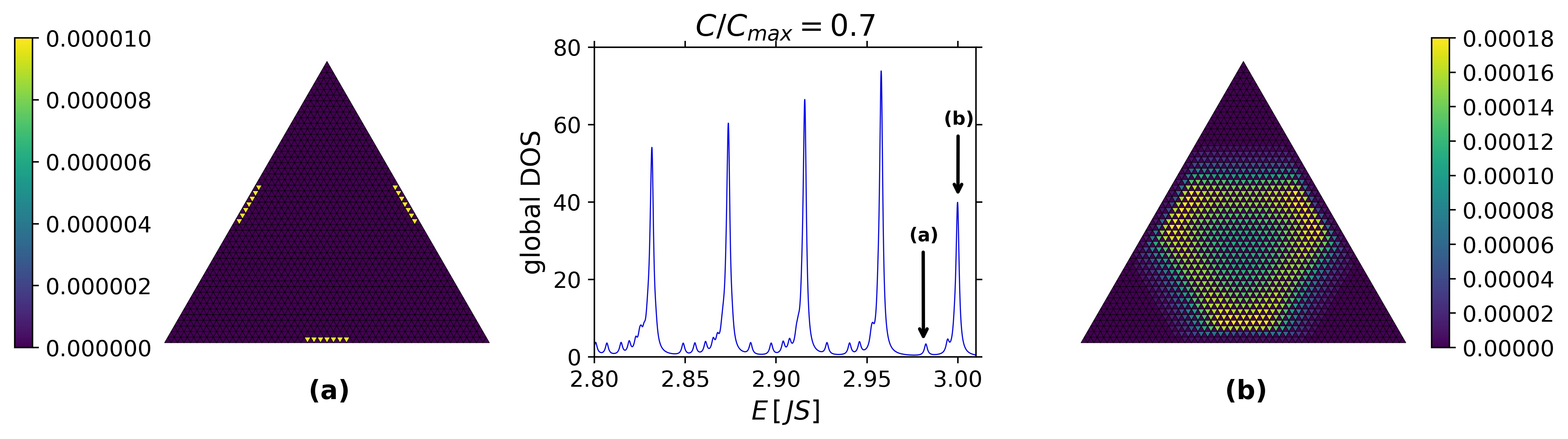}
\caption{
Same as Fig.~\ref{fig:ldosprofile1}, but now for $C/\Cmax=0.7$.}
\label{fig:ldosprofile2}
\end{figure*}

In Fig.~\ref{fig:ldos1} we show results for the local dynamic susceptibility for $\beta=1000$ and different values of the strain $C/\Cmax$, i.e., for the same parameters as in Fig.~3 of the main paper. These figures confirm the existence of equally spaced Landau levels, with a spacing proportional to $C$.
This is illustrated in Fig.~\ref{fig:spacing} which displays the Landau-level spacing $\Delta E$ as function of $C$ for different values of $\beta$. In the limit of large $\beta$, the field-theory prediction \eqref{magen} is fulfilled exactly, while smaller $\beta$ lead to an increased level spacing: The lattice correction $A_{\vec K}$ in Eq.~\eqref{pseudoA} renders the pseudo-magnetic field both inhomogeneous and slightly larger (compared to the $\beta\to\infty$ limit).

The spatial structure of the eigenmodes is shown in Figs.~\ref{fig:ldosprofile1} and \ref{fig:ldosprofile2} which illustrate the spatial distribution of the local susceptibility $\chi_{ii}$ \eqref{chiloc} integrated over a small energy window corresponding to a group of almost degenerate states. While the highly degenerate Landau levels live in the center of the sample, Figs.~\ref{fig:ldosprofile1}(b) and \ref{fig:ldosprofile2}(b), the small-intensity peaks in the DOS correspond to states localized near the sample edge, Figs.~\ref{fig:ldosprofile1}(a) and \ref{fig:ldosprofile2}(a). Consequently, their relative intensity decreases with system size.
We recall that strain-induced pseudo-Landau levels do not display chiral edge states, as the underlying hopping problem preserves time-reversal symmetry. Consequently, any edge states are non-universal, i.e., depend on boundary conditions, sample shape etc., and in particular disappear in the combined limit of $\beta\to\infty$ and $\bar C\to 0$.

\bigskip

\subsection{Protocol for ``perfect strain''}

For completeness, we summarize the model properties in the combined limit of $\beta\to\infty$ and $\bar C\to 0$, as discussed in Refs.~\onlinecite{schomerus14,rachel16b}. In this limit, the lattice distortion is infinitesimal, such that we can consider an effectively undistorted honeycomb spin lattice with inhomogeneous exchange couplings. The latter vary linearly with position according to \eqourjij in the main text. In the limit $C=\Cmax\equiv 1/N$ they are given by\cite{rachel16b}
\begin{equation}
\label{perf-strain}
J^{(N)}_{\vec r,j}=  J \frac{N-1-2\,{\vec r}\cdot{\vec \delta}_j}{N}
\end{equation}
where $\vec r$ denotes the position of a B site, with $\vec r = 0$ defining the center of the system, and $j$ the bond direction. This pattern of bond strengths naturally cuts out a triangle of linear size $N$ from an infinite honeycomb-lattice sheet, see Fig.~1 of Ref.~\onlinecite{schomerus14} and Fig.~S1 of Ref.~\onlinecite{rachel16b}.

Clearly, the ``perfect'' magnon Landau-level spectrum in Fig.~3(c) of the main paper must related to the exact spectrum discovered and derived in Ref.~\onlinecite{rachel16b} for the case of electrons hopping on suitable bipartite lattices; a detailed discussion will be given elsewhere.
